\documentstyle[12pt]{article}

\batchmode
  \newfont{\footscrfont}{rsfs10}
  \newfont{\footbbbfont}{msbm10}
\errorstopmode

\newif\ifscrf\scrftrue
\ifx\footscrfont\nullfont
  \scrffalse
\fi

\newif\ifamsf\amsftrue
\ifx\footbbbfont\nullfont
  \amsffalse
\fi

\def\ppnumber{IASSNS-HEP-94/19}
\def\ppdate{March 1994}
\def\pplogo{\vbox{\kern-\headheight\kern -1pt
\halign{##&##\hfil\cr&{
\ppnumber}\cr\rule{0pt}{2.5ex}&\ppdate\cr}
}}

\makeatletter
\date{}
\def\dedicatory#1{\def\@date{\normalsize\it#1}}
\def\subjclass#1{\def\@thefnmark{}\@footnotetext{1991
    {\it Mathematics Subject Classification.} #1}}
\def\keywords#1{\def\@thefnmark{}\@footnotetext{
    {\it Key words and phrases.} #1}}

\def\ps@firstpage{\ps@empty \def\@oddhead{\hss\pplogo}%
  \let\@evenhead\@oddhead 
}
\def\maketitle{\par
 \begingroup
 \def\thefootnote{\fnsymbol{footnote}}
 \def\@makefnmark{\hbox
 to 0pt{$^{\@thefnmark}$\hss}}
 \if@twocolumn
 \twocolumn[\@maketitle]
 \else \newpage
 \global\@topnum\z@ \@maketitle \fi\thispagestyle{firstpage}\@thanks
 \endgroup
 \setcounter{footnote}{0}
 \let\maketitle\relax
 \let\@maketitle\relax
 \gdef\@thanks{}\gdef\@author{}\gdef\@title{}\let\thanks\relax}

\def\abstract{\if@twocolumn
\section*{Abstract}
\else \small
\begin{center}
{\bf ABSTRACT}
\end{center}
\quotation
\fi}

\newif\iffn\fnfalse

\@ifundefined{reset@font}{\let\reset@font\empty}{} 
\long\def\@footnotetext#1{\insert\footins{\reset@font\footnotesize
    \interlinepenalty\interfootnotelinepenalty
    \splittopskip\footnotesep
    \splitmaxdepth \dp\strutbox \floatingpenalty \@MM
    \hsize\columnwidth \@parboxrestore
   \edef\@currentlabel{\csname p@footnote\endcsname\@thefnmark}\@makefntext
    {\rule{\z@}{\footnotesep}\ignorespaces
      \fntrue#1\fnfalse\strut}}}
\makeatother

\ifamsf
  \newfont{\bigbbbfont}{msbm10 scaled\magstep2}
  \newfont{\bbbfont}{msbm10 scaled\magstep1}  
  \newfont{\smallbbbfont}{msbm8}
  \newfont{\tinybbbfont}{msbm6}
  \newfont{\smallfootbbbfont}{msbm7}
  \newfont{\tinyfootbbbfont}{msbm5}
\fi

\ifscrf
  \newfont{\scrfont}{rsfs10 scaled\magstep1}  
  \newfont{\smallscrfont}{rsfs7}
  \newfont{\tinyscrfont}{rsfs7}
  \newfont{\smallfootscrfont}{rsfs7}
  \newfont{\tinyfootscrfont}{rsfs7}
\fi

\ifamsf
  \newcommand{\Bbb}[1]{\iffn
      \mathchoice{\mbox{\footbbbfont #1}}{\mbox{\footbbbfont #1}}
      {\mbox{\smallfootbbbfont #1}}{\mbox{\tinyfootbbbfont #1}}\else
      \mathchoice{\mbox{\bbbfont #1}}{\mbox{\bbbfont #1}}
      {\mbox{\smallbbbfont #1}}{\mbox{\tinybbbfont #1}}\fi}
\else
  \def\bigbbbfont{\bf}
  \def\Bbb{\bf}
\fi

\ifscrf
  \newcommand{\Scr}[1]{\iffn
    \mathchoice{\mbox{\footscrfont #1}}{\mbox{\footscrfont #1}}
    {\mbox{\smallfootscrfont #1}}{\mbox{\tinyfootscrfont #1}}\else
    \mathchoice{\mbox{\scrfont #1}}{\mbox{\scrfont #1}}
    {\mbox{\smallscrfont #1}}{\mbox{\tinyscrfont #1}}\fi}
\else
  \def\Scr{\cal}
\fi

\def\C{{\Bbb C}}

\def\P{{\Bbb P}}

\def\R{{\Bbb R}}
\def\Z{{\Bbb Z}}

\def\opeq#1{\advance\lineskip#1 \advance\baselineskip#1
	\advance\lineskiplimit#1}
\def\eqalign#1{\null\,\vcenter{\opeq{2.5\jot}\mathsurround=0pt
	\everycr={}\tabskip=0pt
	\halign{\strut\hfil$\displaystyle{##}$&$\displaystyle{{}##}$\hfil
	\crcr#1\crcr}}\,\null}

\def\sm{$\sigma$-model}
\def\nlsm{non-linear \sm}

\def\CY{Calabi-Yau}
\def\LG{Landau-Ginzburg}

\def\cM{{\Scr M}}
\def\cA{{\Scr A}}
\def\cK{{\Scr K}}

\def\cMc{{\hfuzz=100cm\hbox to 0pt{$\;\overline{\phantom{X}}$}\cM}}

\def\ff#1#2{{\textstyle\frac{#1}{#2}}}

\begin{document}
\setcounter{page}0
\title{\LARGE Minimum Distances in Non-Trivial\\
String Target Spaces\\[10mm]
}
\author{\vbox{
\begin{tabular}{c}
\normalsize Paul S. Aspinwall\\[5mm]
\normalsize School of Natural Sciences\\
\normalsize Institute for Advanced Study\\
\normalsize Princeton, NJ  08540
\end{tabular}}
}

{\hfuzz=10cm\maketitle}

\def\Large{\large}
\def\LARGE{\large\bf}

\vskip 3cm		

\begin{abstract}

The idea of minimum distance, familiar from $R\leftrightarrow1/R$
duality when the string target space is a circle, is analyzed for less
trivial geometries. The particular geometry studied is that of a
blown-up quotient singularity within a Calabi-Yau space and mirror
symmetry is used to perform the analysis. It is found that zero
distances can appear but that in many cases this requires other
distances within the same target space to be infinite. In other cases
zero distances can occur without compensating infinite distances.

\end{abstract}

\vfil\break

\section{Introduction}		\label{s:intro}

Questions regarding very small distances in theories of quantum
gravity inevitably run into conceptual difficulties. Classical general
relativity is based on the concept of a space-time metric and so in
this context one can be clear about one's definition of length. In a
theory of quantum gravity this issue is more model dependent. In
models such as string theory, which to date appears to provide by far
the most promising candidate for a consistent theory of quantum
gravity, the metric does not appear as a fundamental concept. In order
to make contact with classical gravity one analyzes the \nlsm\ as an
effective theory for the string theory. One may then retrieve the
classical equations of general relativity \cite{Cal:sm} in some
low-energy limit.

Within the context of string theory therefore the most natural
definition of length would appear to come from the \nlsm. A
conventional description of string theory might be as a theory based
on 1 dimensional strings such that the action is given by the area of
the world-sheet swept out. Our approach will be somewhat the reverse
of this --- one assumes string theory has a more fundamental
definition than above and one defines area to be the object determining
the action. This definition of area and thus distance is what is meant
by saying that the string \nlsm\ is used to define distance.

This approach of using the \sm\ is effectively used in the well-known result
\cite{KY:rr,SS:rr}
concerning a string with a circle of radius $R$ as target space.
Namely, such a theory is equivalent to a string on a circle of radius
$\alpha^\prime/R$. Thus the complete moduli space of strings on circles should
only contain the interval $0\leq R\leq\sqrt{\alpha^\prime}$
or $\sqrt{\alpha^\prime}\leq R\leq\infty$ but not both. Since we
assume we want to make contact with classical physics at distance
scales $\gg\sqrt{\alpha^\prime}$ it is sensible to choose the latter.
Thus we say that string theory ``cuts off'' distances
$<\sqrt{\alpha^\prime}$ since they do not appear anywhere in the
moduli space.

This idea of a minimum distance of order $\sqrt{\alpha^\prime}$ fits
in well with other arguments both within the context of string theory
and otherwise. A direct argument for a minimum distance in string
theory along the
lines of an uncertainty principle was made in \cite{KPP:min}. This was
also in agreement with earlier ideas of \cite{Ven:min} and the results
suggested by scattering strings at high
energies studied in \cite{GM:}. One can also argue for minimum lengths
to appear naturally in other forms of quantum gravity --- see, for
example, \cite{Garay:} for a review of these ideas.

In \cite{AGM:sd} more complicated string target spaces were considered
with the condition of $N$=(2,2) world-sheet supersymmetry imposed. For
a compact smooth target space of definite metric (and vanishing
``torsion'') this implies the \CY\
condition \cite{CHSW:}.
In the case of a circle (or a torus, for which similar results occur)
the minimum distance appears due to non-trivial $\pi_1$, i.e.,
fundamental group, for the target space. The
$R\leftrightarrow\alpha^\prime/R$ symmetry appears due to an
interchange of momentum modes and winding modes. In \cite{AGM:sd}, and
in this paper, the target spaces are simply-connected but
have nontrivial $\pi_2$. Thus it is instantons rather than
solitons which provide the interesting structure.

The approach taken in \cite{AGM:sd} was similar to the above method
for finding the minimum radius of a circle. That is, one constructs
the moduli space such that it contains theories corresponding to large
distances and thus, presumably, classical general relativity and then
one labels each point in the moduli space according to the size of
target space it corresponds to. The minimum size is then the smallest
value thus acquired in the entire moduli space. In the case of the
circle there is only one distance that can be measured, namely the
radius of the circle. In the case of a less trivial target space there
are potentially many sizes corresponding to different parts of the
target space that can be independently varied (at least
classically). In
\cite{AGM:sd} only one size at a time was measured while the other
``independent'' sizes were held at infinity. It was in this context that
some zero distances were measured.

We are therefore now in contradiction with other work on minimum distances.
All of the other works assumed that space looks, to a greater or lesser
degree, locally flat. While this may appear quite reasonable within
the context of classical general relativity, it is not at all clear
that such an assumption is always valid in string theory. In
particular, the perturbative analysis of the \nlsm\ loop by loop gives
conditions for conformal invariance similar to the analysis performed
in \cite{KPP:min}. It is known however that non-perturbative effects
are very important in the \nlsm\ and indeed it is precisely these
effects we are studying in this paper. It should therefore not be a
complete surprise if the results in \cite{AGM:sd} and
this paper are not quite in
agreement with a universal minimum length.

The main purpose of this paper will
be to relax the constraint imposed in \cite{AGM:sd} of holding some
sizes at infinity, i.e., we will ask the question as to
whether zero distances can be measured within a target space of finite
size.

By varying the various sizes with the target space, i.e., by varying
various components of the K\"ahler form, various things might happen
classically to result in some zero distance. Firstly, the ``sides'' of
the target space may be brought together so that the resultant
dimension of the target space decreases. An extreme example of this is
shrinking the entire space down to a point. All the deformations of the
K\"ahler form on a torus resulting in zero size are of this
dimension-lowering type. It was found in \cite{AGM:sd}, at least
within the context of models of the type studied in \cite{Bat:m}, that
string theory imposed a non-zero minimum distance on such
deformations.
Another possibility that can arise as the K\"ahler form is varied
occurs when a subspace of the target space shrinks down to nothing but
the rest of the target space keeps its dimensionality intact. It was
found that in this case, again at least for the class of models
considered, string theory never imposes a nonzero limit on such a process.
Thus it
might appear that the general philosophy of a minimum size or volume might
by true in some form in the context of string theory but that it would
be too na\"\i ve to state it in terms a simply as ``string theory
removes from all consideration distances $<\sqrt{\alpha^\prime}$''.

In this paper we wish to concentrate on a particular example of a
subspace shrinking down. This will consist of some divisor (i.e.,
complex codimension one space) being ``blown-down'' resulting in a
singular target space where the singularity is locally of the form of
a quotient singularity.
String theory on this singular space, known as an ``orbifold'', is fairly
well-understood \cite{DHVW:}.
See, for example, \cite{me:orb2} for a review
of the stringy analysis of such a blowing-down procedure. It is common
in the physics literature to assert that an orbifold can be written in
the form $M/G$ for some smooth manifold $M$ and some finite group $G$.
We will use the more general definition in which the singularities need
only be written {\em locally\/} in the form of a quotient. Since we
are only really interested in the properties of marginal operators in
this paper and since the twisted marginal operators are localized
around the quotient singularities, whether the quotient singularities
are globally quotients should not be particularly important for our
purposes.

As indicated above, the various sizes within a single target space are
classically independent. The question we will be addressing therefore
lies in the realms of quantum geometry. This should come as no
surprise since our problem involves small distances which is where we
expect classical geometry to break down, at least within the context
of string theory. As we shall see the minimum size of one part of
the target space does generally depend on some of the other sizes
within the target.

In section \ref{s:build} we will review how to build examples of the
desired moduli space using mirror symmetry. For this reason our
discussion will allow us to analyze only models with an $N$=(2,2)
superconformal symmetry. We will also review the means by which the
mirror map may be used to label each point in the moduli space by the
size of target space to which it corresponds.

In section \ref{s:hyp} we will perform this labeling by solving a
system of hypergeometric equations. Due to the technicality of this
process we will concentrate on a couple of examples rather than
attempt any general discussion of a solution.
Finally in section \ref{s:int} we will discuss the meaning of the results.


\section{Building the Moduli Space} 		\label{s:build}

In this section we will construct moduli spaces of conformal field
theories. This should be thought of as the stringy analogue of the
moduli space of Ricci-flat metrics for vacuum solutions of classical general
relativity. Little technology currently exists for building the moduli
space of a generic conformal field theory but in the case of $N$=(2,2)
superconformal field theories the situation is far more tractable. If
one further imposes that the central charge have value $c=3d$ for
$d\in\Z$ and that for any NS field in the theory, $Q,\bar Q\in\Z$,
where $Q$ and $\bar Q$ are the left and right-moving $U(1)$ charges
respectively, then there is a general belief that this conformal field
theory should be equivalent to a \nlsm\ with a target space of a
Calabi-Yau $d$-fold or some generalized notion thereof. The most
promising generalized notion is probably the infrared fixed point of
the gauged linear \sm\ with \LG-type potential introduced in
\cite{W:phase}. We will assume that our conformal field theory
corresponds to some \CY\ manifold $X$ or at least belongs to a phase
in a moduli space which also contains a phase for $X$ in the sense
of \cite{W:phase,AGM:II}.

Around any point in the moduli space, the tangent directions are given
by marginal operators. For the theory in questions we can divide these
marginal operators into two groups. One group gives deformations of
complex structure of $X$ and the other gives deformation of the
complexified K\"ahler form of $X$. Recall that the complexification of
the K\"ahler form arises from the general form of the \nlsm:
\begin{equation}
  S = \frac i{4\pi\alpha^\prime}\int\left\{g_{i\bar\jmath}
	(\partial u^i\bar\partial u^{\bar\jmath}+
	\bar\partial u^i\partial u^{\bar\jmath})
	-iB_{i\bar\jmath}(\partial u^i\bar\partial u^{\bar\jmath}-
	\bar\partial u^i\partial u^{\bar\jmath})
	\right\}\,d^2z,         \label{eq:sm}
\end{equation}
The field theory based on this action depends only the cohomology
class of
the real two-form given by $B_{i\bar\jmath}$.
Assuming $h^{2,0}(X)=0$, we construct $B+iJ\in H^2(X,\C)$ as the
complexified K\"ahler form where $J$ is the classical K\"ahler form
derived from $g_{i\bar\jmath}$. An important observation is that,
semi-classically, the
field theory also obeys the symmetry
\begin{equation}
  B+iJ\cong B+iJ+4\pi^2\alpha^\prime L,\quad\forall L\in H^2(X,\Z).
\end{equation}

The deformations of complex structure have no quantum corrections.
That is, the marginal operators corresponding to such deformations of
the conformal field theory may be integrated along to build up a
moduli space which is isomorphic to the classical moduli space of
complex structures of $X$. The other marginal operators are different
however. Instantons affect the form of this moduli space so that the
classical moduli space and conformal field theory moduli space differ.
See, for example, \cite{W:AB} for an account of the way instanton
affects appear in this context.
The classical moduli space for $B+iJ$ would be of the form $\cK\times
(S^1)^{h^{1,1}}$ where $\cK$ is the classical moduli space of $J$,
i.e., the K\"ahler cone. The cone-like structure appears since if $J$
is a valid K\"ahler form on $X$ then so is $\lambda J$ for $\lambda$ a
positive real number. It is precisely this structure which should be
modified if we expect minimum distances to appear --- if $J$ is a
valid K\"ahler form for the string target space it should not
necessarily follow that $\lambda J$ is too.

The trick which allows us to construct the entire conformal field
theory moduli space is that provided by mirror symmetry \cite{GP:orb}.
Namely there may be another space $Y$ on which a \nlsm\ produces the
same conformal field theory as that on $X$ except that the r\^ole of
the marginal operators are interchanged: those giving deformations of
complex structure on $Y$ give deformations of K\"ahler form on $X$ and
{\em visa versa}. Thus to build the stringy version of the moduli
space of K\"ahler forms on $X$ we simply build the moduli space of
complex structures on $Y$ and assert it is isomorphic to the desired
moduli space.

As an example we will consider the \CY\ manifold studied at some length
in \cite{CDFKM:I}. Let the space $\P_{\{2,2,2,1,1\}}^4$ be defined as
$(\C^5-\{0\})/\C^*$ where the coordinates of $\C^5$ are $(X_0,X_1,
\ldots, X_5)$ and the $\C^*$-action is given by
\begin{equation}
  (X_0,X_1,X_2,X_3,X_4)\mapsto(\lambda^2X_0,\lambda^2X_1,
  \lambda^2X_2,\lambda X_3,\lambda X_4),\qquad\lambda\in\C^*.
\end{equation}
This space has a quotient singularity locally of the form $\C^2/\Z_2$
along the $\P^2$, $X_3=X_4=0$. This quotient singularity may be blown
up with $\P^1\times\P^2$ as exceptional divisor. Let $X$ be the
resolved hypersurface
\begin{equation}
  X_0^4+X_1^4+X_2^4+X_3^8+X_4^8=0
\end{equation}
within the resolved $\P_{\{2,2,2,1,1\}}^4$.
$X$ has $h^{1,1}=2$, with one component given by the ambient
$\P_{\{2,2,2,1,1\}}^4$ and the other given by the exceptional divisor in
the blow-up.
The method of \cite{GP:orb} tells us how to construct the mirror, $Y$,
as a blown-up orbifold of this. In order to study the moduli space
$\cM$ of complex structures of $Y$, we need not concern ourselves with
these latter blow-up modes. Instead we write down the most general
form of defining equation for $Y$ compatible with the orbifolding.
This is
\begin{equation}
W=a_0X_0X_1X_2X_3X_4+a_1X_3^4X_4^4+a_2X_0^4+a_3X_1^4
	+a_4X_2^4+a_5X_3^8+a_6X_4^8,  	\label{eq:con1}
\end{equation}
where $a_j$ are arbitrary complex numbers. Rescaling the $X_i$
coordinates gives a 5-dimensional space of reparametrizations of this
equation thus leaving the $a_j$'s to span a 2-dimensional moduli space
in agreement with $h^{1,1}(X)=2$. Actually, as explained in
\cite{AGM:II} this gives the moduli space the structure of a toric
variety. This toric information also allows us to build the natural
compactification, $\cMc$, of $\cM$ within the context of mirror symmetry by
using the ``secondary fan''. This procedure requires a fairly lengthy
explanation and the reader is encouraged to consult \cite{AGM:II} for
a full account of this process.

The general result required in this context is as follows. Let there
be $n$ ``homogeneous'' coordinates $X_i$, $i=0,\ldots, n-1$ and let
there be $N$ terms in the general defining equation for $Y$ with
coefficients $a_j$, $j=0,\ldots, N-1$. The moduli space can then be
built as a toric variety as a compactification:
\begin{equation}
  \cMc \supset \frac{(\C^*)^N}{(\C^*)^n}\cong(\C^*)^{N-n}.
\end{equation}
The dimension of the moduli space will thus be $N-n$. This may be
equal to, or less than $h^{1,1}(X)$. The latter case arises when
independent elements of homology of $X$ cannot be written in terms of
independent elements of homology of the ambient space. In this case we
will not be studying the entire moduli space of complexified K\"ahler
forms of $X$ but only the ``toric part''. The compact space $\cMc$ is
described by coordinate patches each having coordinates $z_l$,
$l=1,\ldots, N-n$ given by some $(\C^*)^n$-invariant product (or
quotient) of the $a_j$'s. The precise $(\C^*)^n$-invariant product
used for each coordinate in each patch dictates how the patches are sewn
together and is given by the data of the secondary fan --- one cone
for each coordinate patch.

Let us now try to identify each point in $\cMc$ with a particular $X$.
Each origin of each coordinate patch marks a special point in $\cMc$
called a ``limit point''. The number of limit points is thus equal to
the number of coordinate patches used to build $\cMc$
which is equal to the number of cones
in the secondary fan. The work of \cite{W:phase} and
\cite{AGM:II,AGM:I,AGM:mdmm} may then be used to identify one of these limit
points with the conformal field theory associated to the \nlsm\
defined on the large radius limit of $X$. To be more precise, let us
expand the K\"ahler form
\begin{equation}
  B+iJ=\sum_l(B+iJ)_le_l,
\end{equation}
where $e_l$ are positive generators of
$H^2(X,4\pi^2\alpha^\prime \Z)$. That is, $\int_C e_l$, is positive
for any (one complex dimensional) curve in $X$. It is then claimed
\cite{AGM:mdmm} that one of the coordinate patches within $\cMc$ given by
$z_l^{(0)}$ specifies $X$ as
\begin{equation}
  (B+iJ)_l = \frac1{2\pi i}\log z_l^{(0)} +
O(z_1^{(0)},z_2^{(0)},\ldots),	\label{eq:asym}
\end{equation}
near the origin $z_l^{(0)}=0$, $l=1,\ldots, N-n$. The origin itself
thus corresponds to the limit of $X$ with all components of
$J\to\infty$. This is the ``large radius limit''.

By making this identification we have thus ensured that large
distances are included in our moduli space. In some respects this is
similar to the step of choosing the interval $\sqrt{\alpha^\prime}\leq
R\leq\infty$ rather than $0\leq R\leq\sqrt{\alpha^\prime}$ in the
definition of the stringy moduli space of a circle.

Now all we need to do is to extend the definition of $(B+iJ)_l$ away
from the large radius limit over the whole of $\cMc$. To attempt to do
this we use the result conjectured in \cite{CDGP:} and proven in
\cite{BCOV:big}. This result states that local geometry of the moduli
space of $N$=2 theories dictates that there exist elements of homology
$\gamma_0$ and $\gamma_l$ on $Y$ such that
\begin{equation}
  (B+iJ)_l = \frac{\displaystyle\int_{\gamma_l}\Omega}
	{\displaystyle\int_{\gamma_0}\Omega}, \label{eq:period}
\end{equation}
where $\Omega$ is a highest holomorphic form on $Y$. The choice of
which cycles are used as $\gamma_0$ and $\gamma_l$ is governed by the
global geometry of the moduli space as discussed in \cite{Mor:cp}.
The periods in (\ref{eq:period}) (i.e., the numerator and denominator
in the right hand side) may be evaluated as in \cite{lots:per}. In
this paper, as in \cite{AGM:sd}, we will proceed in a slightly
indirect manner. We find some differential
equations satisfied by $\int_\gamma\Omega$
and then use (\ref{eq:asym}) to choose
which solutions we require. This system of differential equations is
known as the ``hypergeometric system'' to which we now turn our
attention.


\section{The Hypergeometric System}		\label{s:hyp}

Following the construction of \cite{Bat:m}, a hypersurface in a
$(n-1)$-dimensional
toric variety may be represented by a set of $N$ points $\cA$ in a
hyperplane in $\R^n$. Each point in $\cA$ may be thought of as
representing each monomial in some constraint such as (\ref{eq:con1})
where the more generalized notion of homogeneous coordinate
\cite{Cox:} may be used.
The coordinates of each point are given by the degrees to which each
homogeneous coordinate is raised. A lattice structure is also imposed
within the space $\R^n$ such that the point set $\cA$ is the
intersection of this lattice with some convex polytope. Let us now
define new coordinates based on this lattice and let
$\alpha_j\in\cA\;$ have coordinates $\alpha_{ji}$ for $i=0,\ldots,n-1$.
The hypergeometric system \cite{GZK:h} is formed by the set of differential
operators
\begin{equation}
\eqalign{
  Z_i &= \left(\sum_{j=0}^{N-1}
	\alpha_{ji} a_j\frac\partial{\partial a_j}\right)
	-\beta_i\cr
  \Box_l &= \prod_{m_{lj}>0}\left(\frac\partial{\partial
	a_j}\right)^{m_{lj}} - \prod_{m_{lj}<0}\left(\frac\partial{\partial
	a_j}\right)^{-m_{lj}},\cr}	\label{eq:ZBox}
\end{equation}
where $i=0,\ldots,n-1$ and $l=1,\ldots,N-n$ labels a relationship
\begin{equation}
  \sum_{j=0}^{N-1}m_{lj}\alpha_{ji}=0, \qquad\forall i. \label{eq:cnd1}
\end{equation}
We may tie the labeling of these relationships in with the coordinate
patches on $\cMc$ we used in section \ref{s:build}. If we write
\begin{equation}
  z_l = \prod_{j=0}^{N-1} a_j^{m_{lj}} \label{eq:z2a}
\end{equation}
then the $(\C^*)^n$ invariance of $z_l$ is equivalent to the condition
(\ref{eq:cnd1}).
Now we look for a function $\Phi(a_0,a_1,\ldots,a_{N-1})$ such that
\begin{equation}
  Z_i \Phi = \Box_l\Phi = 0, \qquad\forall i,l.  \label{eq:PF}
\end{equation}
Let us further require that our coordinates for the lattice are chosen
such that the hyperplane condition is imposed by $\alpha_{j0}=1$ for
any $j$ and let the unique point \cite{Bat:m} within the interior of
the polytope be $\alpha_0$ and have coordinates $\alpha_{0i}=0$ for
$i>0$. It was then shown in \cite{Bat:var} that the periods
$\int_\gamma\Omega$ satisfy (\ref{eq:PF}) for $\beta_0=-1$ and
$\beta_i=0$ for $i>0$.

It was shown in \cite{GZK:h} that a solution of (\ref{eq:PF}) may be
written
\begin{equation}
  \Phi_\gamma=\left(\prod_{j=0}^{N-1}a_j^{\gamma_j}\right)
  \sum_{\{P_l\}\in\Z^{N-n}}
  \frac{\displaystyle\prod_{l=1}^{N-n}z_l^{P_l}}
  {\displaystyle\prod_{j=0}^{N-1}\Gamma\left(
  \sum_{l=1}^{N-n}m_{lj}P_l+\gamma_j+1\right)},		\label{eq:Phi}
\end{equation}
where $\gamma$ is a vector in $\R^N$ such that
\begin{equation}
  \sum_{j=0}^{N-1}\gamma_j\alpha_{ji}=\beta_i.
\end{equation}
In fact it is a simple matter to show that (\ref{eq:Phi}) formally
satisfies (\ref{eq:PF}). Treating $\Phi_\gamma$ as a sum of terms over
the lattice $\Z^{N-n}$, one can show that $Z_i\Phi_\gamma$ vanishes
individually at each lattice site. The two terms in $\Box_l$ in
(\ref{eq:ZBox}) act on $\Phi_\gamma$ to produce the same expression
except shifted by some lattice vector. Thus the sum over the whole
lattice vanishes. The only property of the gamma function used to show
this result is $\Gamma(x+1)=x\Gamma(x)$.

It is clear that for a generic $\gamma$, the expression (\ref{eq:Phi})
will have zero radius of convergence for $z_l$ and is thus somewhat
useless. If $\gamma$ is tuned to the right value however, the fact
that $\Gamma(x)$ acquires poles for $x=0,-1,-2,-3,\ldots$ may be used to
reduce the sum over all $P_l$ to only $P_l\geq0$. One would then
expect (\ref{eq:Phi}) to converge for sufficiently small $|z_l|$. In
the generic case this is how one generates all the generalized
hypergeometric series which solve the hypergeometric system, i.e., all
solutions may be written in the form $\Phi_\gamma$ for a suitable set
of $\gamma$.

If $\alpha_{ji}$ and $\beta_i$ satisfy certain ``resonance''
conditions \cite{GZK:h} then not all the solutions of the
hypergeometric system may be written in the form (\ref{eq:Phi}). As
luck would have it, for the case we are considering, these parameters
resonate sufficiently badly that {\em none\/} of the solutions may be
written in this form. As we now explain however we may still use the
expression (\ref{eq:Phi}) as a starting point for writing down all the
required solutions. This issue was also studied in \cite{HKTY:}. Our
method is similar but
will use Barnes-type integrals which directly allow the solutions
to be analytically continued.

Firstly note that the expression
\begin{equation}
  \Gamma(x)\Gamma(1-x)=\frac\pi{\sin(\pi x)},
\end{equation}
may be used to take gamma functions from the denominator into the
numerator in (\ref{eq:Phi}). Suppose $M$ is an integer $\geq0$. One
may then write
\begin{equation}
  \frac1{\Gamma(-M+\epsilon)}\simeq\Gamma(M+1)(-1)^M\epsilon,
		\label{eq:denum}
\end{equation}
for small $\epsilon$. Ignoring for a moment the $\epsilon$ factor on the
right-hand side, in the limit $\epsilon\to0$, this identity may be
used to remove some poles from the denominator in (\ref{eq:Phi}). One
can argue that the $\epsilon$ factor on the right side of
(\ref{eq:denum}) contributes only an overall scale to the solution and
can thus be ignored, even though it's zero! If the reader is skeptical
of this manipulation, they should substitute the expression thus obtained
back into the hypergeometric system to verify that it is indeed a
solution.

Let us now illustrate this procedure by applying it to the example
in (\ref{eq:con1}). Denoting $z_1$ and $z_2$ by $x$ and $y$ for
simplicity, analysis of the secondary fan shows that
\begin{equation}
x=\frac{a_5a_6}{a_1^2},\qquad
y=\frac{a_1a_2a_3a_4}{a_0^4}
\end{equation}
are the coordinates whose origin corresponds to the large radius
limit. Now, if we use the value $\gamma=(-1,0,0,\ldots)$ in
(\ref{eq:Phi}) we have
\begin{equation}
a_0^{-1}\sum_{M,N}
  \frac{x^My^N}{\Gamma(-4N)\Gamma^2(M+1)
  \Gamma^3(N+1)\Gamma(N-2M+1)}.
\end{equation}
This is zero since every term in the sum vanishes. Using the above
trick however we may move $\Gamma(-4N)$ up into the numerator giving
\begin{equation}
\Phi_0=a_0^{-1}\!\!\sum_{{N\geq0\atop 0\leq M\leq N/2}}
  \frac{\Gamma(4N+1)\,x^My^N}{\Gamma^2(M+1)
  \Gamma^3(N+1)\Gamma(N-2M+1)}.  \label{eq:bpL}
\end{equation}

This is the unique solution (up to an overall scale) of the
hypergeometric system which is regular at $(x,y)=(0,0)$.
No other value of $\gamma$ gives a convergent series and so we must
obtain all our solutions from this single value of $\gamma$.
We may write (\ref{eq:bpL}) as a Barnes integral:
\begin{equation}
\Phi_0=a_0^{-1}\sum_{M\geq0}
  \frac{x^M}{\Gamma^2(M+1)}\;
  \frac1{2\pi i}\int_C\frac{\Gamma(4s+1)\Gamma(2M-s)}
  {\Gamma^3(s+1)}(e^{-\pi i}y)^s\,ds,  \label{eq:bpLB}
\end{equation}
where $C$ is a contour running to the left of Re$(s)=0$ as shown below
(the poles are shown for the $M=0$ term --- more generally
poles will not occur from $s=0,\ldots,2M-1$):
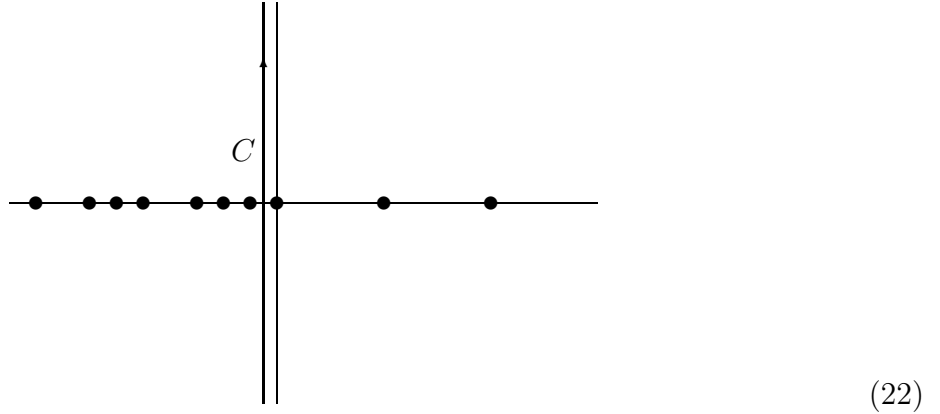
\begin{equation}
\setlength{\unitlength}{0.007in}%
\begin{picture}(440,300)(100,410)
\thinlines
\multiput(300,560)(80,0){3}{\circle*{10}}
\multiput(240,560)(20,0){3}{\circle*{10}}
\multiput(160,560)(20,0){3}{\circle*{10}}
\put(120,560){\circle*{10}}
\put(290,710){\line( 0,-1){300}}
\put(300,710){\line( 0,-1){300}}
\put(100,560){\line( 1, 0){440}}
\put(290,630){\vector(0,1){40}}
\put(275,600){\makebox(0,0){$C$}}
\end{picture}
\end{equation}
We will discuss conditions for convergence of this integral below. One
may think of obtaining (\ref{eq:bpLB}) from (\ref{eq:bpL}) by moving
$\Gamma(N-2M+1)$ up from the denominator into the numerator. This
introduces poles which are rendered finite by the contour integral.
The contour is chosen so that it may be completed to the right to
enclose the correct values of $s$ to contribute to the sum.
We may use this idea further to generate more solutions.

If we further move a $\Gamma(N+1)$ from the denominator we obtain
\begin{equation}
\Phi_y=-a_0^{-1}\sum_{M\geq0}\frac{x^M}{\Gamma^2(M+1)}\;\frac1{2\pi i}
\int_C\frac{\Gamma(4s+1)\Gamma(-s)\Gamma(2M-s)}{\Gamma^2(s+1)
}y^s\,ds,
\end{equation}
along the same contour.
Enclosing the contour to the right and taking residues we obtain
\begin{equation}
\eqalign{
  \Phi_y=a_0^{-1}\sum_{N\geq0}\Biggl\{\sum_{M\leq N/2}
\frac{\Gamma(4N+1)\,x^My^N}
{\Gamma^2(M+1)\Gamma^3(N+1)\Gamma(N-2M+1)}&\biggl[\log y\cr
{}+4\Psi(4N+1)-3\Psi(N+1)&-\Psi(N-2M+1)\biggr]\cr
-\sum_{M>N/2}\frac{\Gamma(4N+1)\Gamma(2M-N)\,x^M(-y)^N}{\Gamma^2(M+1)
\Gamma^3(N+1)}&\Biggr\}.\cr}		\label{eq:BJ1y}
\end{equation}
This is also a solution of the hypergeometric system. The appearance
of the logarithm is precisely what we required from (\ref{eq:asym}).
In fact (\ref{eq:asym}), (\ref{eq:period}) and the behaviour of all the
other solutions around the origin allow us to specify
{\em uniquely\/} that
\begin{equation}
  (B+iJ)_y = \frac1{2\pi i}\frac{\Phi_y}{\Phi_0}.
\end{equation}

The other component $(B+iJ)_x$ may be obtained similarly although this
time more care must be taken to enclose the required poles with the
contour. Let us
consider
\begin{equation}
\Phi_x=a_0^{-1}\sum_{N\geq0}\frac{\Gamma(4N+1)\;(-y)^N}{\Gamma^3(N+1)}\;
 \frac2{2\pi i}\int_C
 \frac{\Gamma(2s-N)\Gamma(-s)}{\Gamma(s+1)}
 (e^{-\pi i}x)^s\,ds,	\label{eq:BJ1xI}
\end{equation}
with the contour given by (e.g., $N=4$)
\begin{equation}
\setlength{\unitlength}{0.007in}%
\begin{picture}(440,300)(100,410)
\thinlines
\put(300,560){\oval( 20, 20)[bl]}
\put(300,560){\oval( 20, 20)[br]}
\put(340,560){\oval( 20, 20)[bl]}
\put(340,560){\oval( 20, 20)[br]}
\put(320,560){\oval( 20, 20)[tr]}
\put(320,560){\oval( 20, 20)[tl]}
\put(360,560){\oval( 20, 20)[tr]}
\put(360,560){\oval( 20, 20)[tl]}
\put(280,560){\circle*{10}}
\put(300,560){\circle*{10}}
\put(320,560){\circle*{10}}
\put(340,560){\circle*{10}}
\put(360,560){\circle*{10}}
\put(380,560){\circle*{10}}
\put(420,560){\circle*{10}}
\put(460,560){\circle*{10}}
\put(500,560){\circle*{10}}
\put(240,560){\circle*{10}}
\put(200,560){\circle*{10}}
\put(160,560){\circle*{10}}
\put(120,560){\circle*{10}}
\put(300,710){\line( 0,-1){300}}
\put(100,560){\line( 1, 0){440}}
\put(290,710){\line( 0,-1){150}}
\put(370,560){\line( 0,-1){150}}
\put(370,430){\vector( 0, 1){ 40}}
\put(290,640){\vector( 0, 1){ 40}}
\put(390,485){\makebox(0,0){$C$}}
\end{picture}
\end{equation}
to yield
\begin{equation}
\eqalign{
\Phi_x=a_0^{-1}\sum_{N\geq0}\Biggl\{\sum_{M\leq N/2}
 &\frac{\Gamma(4N+1)\,x^My^N}{\Gamma^2(M+1)\Gamma^3(N+1)\Gamma(N-2M+1)}
\biggl[\log x-\pi i\cr
 &\qquad\qquad+2\Psi(N-2M+1)-2\Psi(M+1)\biggr]\cr
 &+2\!\!\sum_{M>N/2}\frac{\Gamma(4N+1)\Gamma(2M-N)\,x^M(-y)^N}{\Gamma^2(M+1)
\Gamma^3(N+1)}\Biggr\}.\cr}	\label{eq:BJ1x}
\end{equation}
We then see that
\begin{equation}
  (B+iJ)_x = \frac1{2\pi i}\left(\frac{\Phi_x}{\Phi_0}+\pi i\right).
\end{equation}

We have therefore found expressions giving the exact value of $B+iJ$
at points in the moduli space $\cMc$. Unfortunately the expressions
(\ref{eq:BJ1y}) and (\ref{eq:BJ1x}) only lead to convergent series for
sufficiently small $|x|$ and $|y|$ and cannot be used for the whole of
$\cMc$. In particular they are not valid for areas of $\cMc$ where we
would expect to find minimum distances. In order to find these minimum
distances we analytically continue these hypergeometric functions and
since we have written all the relevant expressions in terms of Barnes
integrals this is a simple matter. There is an issue that we have not
paid sufficient attention to yet which is of great importance, namely
that of branch cuts. In order to analytically continue our functions
over $\cMc$ we are required to specify the branch cuts --- different
branch cuts lead to different values of $B+iJ$ associated to each
point. Indeed, this property may lead one to assert that $B+iJ$ does
not make much sense outside the region of convergence of the above
series. It turns out however that at least some branch cuts may be
made naturally leading to a natural concept of $B+iJ$ away from this
region. These are the branch cuts which have already been made
implicitly above.

The identification $(B+iJ)_l\cong (B+iJ)_l+1$ requires us to make some
choice of interval for the $B$-field. We will cut so that $0<B_l<1$
(note that the cut $-1<B<0$ was implicitly made in \cite{AGM:sd,CDGP:}
hence some minor differences in our formul\ae\ here). This imposes
\begin{equation}
0<\arg\left\{{x\atop y}\right\}<2\pi.\label{eq:xyr}
\end{equation}
All of the above contour integrals are convergent for sufficiently
small $|x|$ and $|y|$ and the condition (\ref{eq:xyr}). Assuming this
is the only branch cut in the neighbourhood of the large radius limit,
this fixes the cut. We analytically continue the above periods simply
by completing the contour to the left rather than the right.

The region of $\cMc$ that we are interested in probing is where zero
distances appear. This happens for the exceptional divisor that grew
out of the $\Z_2$-quotient singularity in the ambient space. The phase
of $\cMc$ we wish to study is the ``orbifold phase''. That is, where
the limit point corresponds to $X$ infinitely large with the $\Z_2$
singularity unresolved. The method of \cite{AGM:II,AGM:sd} tells us that the
patches in this region are given by
\begin{equation}
  \xi=\frac{a_1^2}{a_5a_6}=x^{-1},\qquad\eta=
	\frac{a_2^2a_3^2a_4^2a_5a_6}{a_0^8}=xy^2.  \label{eq:cov1}
\end{equation}

The period $\Phi_0$ actually does not require analytical continuation,
a change of variable gives
\begin{equation}
\Phi_0=a_0^{-1}\sum_{{P,Q\in\Z/2\geq0\atop Q-P\in\Z\geq0}}
  \frac{\Gamma(8Q+1)\,\xi^P\eta^Q}{\Gamma^3(2Q+1)\Gamma(2P+1)
  \Gamma^2(Q-P+1)}.
\end{equation}
If we enclose the contour to the left in (\ref{eq:BJ1xI}) and take
residues we obtain
\begin{equation}
\eqalign{
\Phi_x=-\pi ia_0^{-1}\Biggl\{\sqrt{\xi}\sum_{M,N\geq0}
 &\frac{\Gamma(8N+1)\,\xi^M\eta^N}{\Gamma^2(N-M+\ff12)\Gamma^3(2N+1)
 \Gamma(2M+2)}\cr
 &\qquad+\sqrt{\eta}\sum_{M,N\geq0}
 \frac{\Gamma(8N+5)\,\xi^M\eta^N}{\Gamma^2(N-M+\ff32)\Gamma^3(2N+2)
 \Gamma(2M+1)}\Biggr\}.\cr}
	\label{eq:BJx1o}
\end{equation}
We may also write
\begin{equation}
 \eqalign{\Phi_x+2\Phi_y&=
  a_0^{-1}\sum_{{N\geq0\atop 0\leq M\leq N/2}}
  \frac{\Gamma(4N+1)\,x^My^N}{\Gamma^2(M+1)
  \Gamma^3(N+1)\Gamma(N-2M+1)}\biggl[\log(xy^2)-\pi i\cr
  &\qquad\qquad\qquad{}+8\Psi(4N+1)-6\Psi(N+1)-2\Psi(M+1)\biggr]\cr
  &=a_0^{-1}\sum_{{P,Q\in\Z/2\geq0\atop Q-P\in\Z\geq0}}
  \frac{\Gamma(8Q+1)\,\xi^P\eta^Q}{\Gamma^3(2Q+1)\Gamma(2P+1)
  \Gamma^2(Q-P+1)}\biggl[\log\eta-\pi i\cr
  &\qquad\qquad\qquad{}+8\Psi(8Q+1)-6\Psi(2Q+1)-2\Psi(Q-P+1)\biggr].\cr}
\end{equation}

Now we need to discuss branch cuts in this orbifold phase. There
are branch points at the boundary of the region of convergence of the
periods in the orbifold phase just as in the \CY\ phase above. Thus
the arguments of $\xi$ and $\eta$ should only be allowed to span an
interval of $2\pi$. It is easy to see that $-2\pi<\arg\xi<0$ from
(\ref{eq:cov1}). For $\eta$ we need to look at the discriminant locus
which should be considered as a locus of branch points in some sense.
One component of the discriminant is given by \cite{CDFKM:I} as
\begin{equation}
\eqalign{(4^4y-1)^2&=4^9xy^2\cr
\hbox{i.e.,}\quad(4^4\sqrt{\eta\xi}-1)^2&=4^9\eta.\cr}
\end{equation}
Thus for $x=0$ we have the branch point at $y=4^{-4}$. As $x$
increases from zero this branch point will divide into two. The region
of allowed $\arg(y)$ will thus get squeezed into a smaller region
between the branch cuts. Let $x=Re^{\pi i}$ for a real positive number
$R$ and let $R\to\infty$. One may follow the allowed region of
$\arg(y)$ to avoid the discriminant to find that
$\pi/2<\arg(y)<3\pi/2$. Applying $\eta=xy^2$ we find
\begin{equation}
\eqalign{-2\pi<&\arg\xi<0\cr 2\pi<&\arg\eta<4\pi.\cr}
	\label{eq:xer}
\end{equation}
Actually a simple guiding principle may be used for the general case.
The discriminant locus is described by a large polynomial with integer
(and thus real) coefficients. Thus if we originally have an allowed
region of $0<\arg(z_l)<2\pi$ close to the limit point then as we
switch on values of $z_k=R_ke^{\pi i}$ for the other parameters, the
region of allowed $\arg(z_l)$ must be squeezed symmetrically around
$\arg(z_l)=\pi$. Thus, if any region of the origin of one phase
between cuts can be followed into a neighbouring region, it contains
the direction $\arg(z_l)=\pi$ for all $l$. We substitute this value
into the change of variables from one phase to another to find the
central value of $\arg(z_l^\prime)$ for the other phase.

Before discussing the interpretation of the above results it will be
helpful to have another example in mind. For this we will look at the
five parameter moduli space studied in \cite{AGM:I,AGM:II,AGM:sd}.
In this case $X$ is the resolution of a hypersurface in
$\P_{\{6,6,3,2,1\}}^4$ and $Y$ has general defining equation
\begin{equation}
\eqalign{a_0 X_0X_1X_2X_3X_4 +
	a_1 X_2^3X_4^9 &+ a_2 X_3^6X_4^6 + a_3 X_3^3X_4^{12} +
	a_4 X_2^3X_3^3X_4^3\cr
	&+a_5X_0^3+a_6X_1^3+a_7X_2^6+a_8X_3^9+a_9X_4^{18} = 0.\cr}
	\label{eq:gen}
\end{equation}
This has a remarkably rich phase structure \cite{AGM:II}. The behaviour
we are interested in concerns passage between \CY\ phases and orbifold
phases. The quotient singularity in question will be an isolated
$\Z_3$ singularity. For the \CY\ phase we have coordinates
\begin{equation}
\eqalign{z_1&=\frac{a_1a_3a_5a_6}{a_0^3a_9}\cr
	z_2&=\frac{a_2a_9}{a_3^2}\cr
	z_3&=\frac{a_3a_8}{a_2^2}\cr
	x&=-\frac{a_1a_7a_8}{a_4^3}\cr
	y&=\frac{a_2a_4}{a_1a_8}.\cr}
\end{equation}
(See \cite{AGM:sd} for a discussion of the sign definition appearing
above for $x$.) The period regular at the origin may then be computed as
\begin{equation}
\Phi_0=\sum_{N,M,P_i\geq0}\frac{\Gamma(3P_1+1)\;(-z_1)^{P_1}
z_2^{P_2}z_3^{P_3}(-x)^My^N}
{\left\{\begin{array}{c}
\Gamma(P_1+M-N+1)\Gamma(P_2-2P_3+N+1)\Gamma(P_1-2P_2+P_3+1)\\
\times\Gamma(N-3M+1)\Gamma^2(P_1+1)\Gamma(M+1)\\
\times\Gamma(P_3+M-N+1)\Gamma(P_2-P_1+1)
\end{array}\right\}}.
\end{equation}
In the above and from now on, we will drop the $a_0^{-1}$ factor common
to all periods.

The orbifold region is given by coordinates $z_1,z_2,z_3,\xi$
and $\eta$ where
\begin{equation}
\eqalign{\xi&=-\frac{a_4^3}{a_1a_7a_8}\cr
	\eta&=\frac{a_2^3a_7}{a_1^2a_8^2}\cr}
\end{equation}
For simplicity let us set $z_1=z_2=z_3=0$. This implies $\Phi_0=1$.
We also impose the restrictions of (\ref{eq:xyr}). Choosing phases by
$\xi=x^{-1}$ and $\eta=e^{-\pi i}xy^3$ and finding the allowed region
of $\arg(y)$ from the above reasoning we obtain (\ref{eq:xer}) again.
Now we have
\begin{equation}
  \Phi_x=\sum_{N\geq0}\frac{(-y)^N}{\Gamma(N+1)}\;
 \frac{3}{2\pi i}\int_C\frac{\Gamma(3s-N)\Gamma(-s)}
 {\Gamma^2(s-N+1)}(e^{-\pi i}x)^s\,ds,
\end{equation}
with $C$ just to the left of Re$(s)=N$. This gives
\begin{equation}
  \Phi_x=\log x-\pi i+3\sum_{\hbox{\hidewidth\scriptsize\begin{tabular}{c}
  $N\geq0$\\$M\geq N$\\
  except $N=M=0$\end{tabular}\hidewidth}}\;
  \frac{\Gamma(3M-N)}{\Gamma(M+1)\Gamma(N+1)\Gamma^2(M-N+1)}
  x^M(-y)^N.
\end{equation}
Completing to the left we obtain, to leading order
\begin{equation}
  \Phi_x=-\frac{\Gamma(\ff13)}{\Gamma^2(\ff23)}e^{\frac{\pi i}3}
  \xi^{\frac13}+3\frac{\Gamma(\ff23)}{\Gamma^2(\ff13)}
  \eta^{\frac13}+\ldots		\label{eq:BJ2xo}
\end{equation}

For the other period we use
\begin{equation}
  \Phi_y=\sum_{M\geq0}\frac{(-x)^M}{\Gamma(M+1)}\frac1{2\pi i}
  \int_C\frac{\Gamma^2(s-M)}{\Gamma(s+1)\Gamma(s-3M+1)}y^s\,ds,
\end{equation}
where $C$ is given by (e.g., $M=4$)
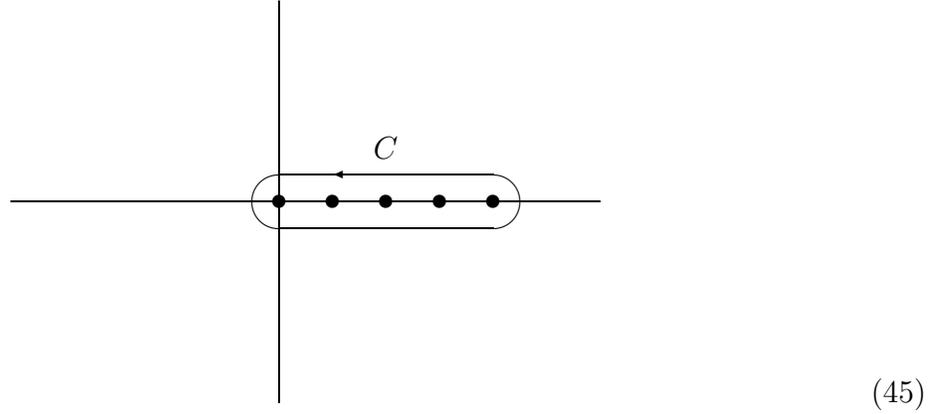
\begin{equation}
\setlength{\unitlength}{0.007in}%
\begin{picture}(440,300)(100,410)
\thinlines
\put(300,560){\circle*{10}}
\put(340,560){\circle*{10}}
\put(380,560){\circle*{10}}
\put(420,560){\circle*{10}}
\put(460,560){\circle*{10}}
\put(380,560){\oval(200,40)}
\put(300,710){\line( 0,-1){300}}
\put(100,560){\line( 1, 0){440}}
\put(380,580){\vector(-1,0){40}}
\put(380,600){\makebox(0,0){$C$}}
\end{picture}
\end{equation}
This gives
\begin{equation}
  \Phi_y=\log y-\sum_{\hbox{\hidewidth\scriptsize\begin{tabular}{c}
  $N\geq0$\\$M\geq N$\\
  except $N=M=0$\end{tabular}\hidewidth}}\;
  \frac{\Gamma(3M-N)}{\Gamma(M+1)\Gamma(N+1)\Gamma^2(M-N+1)}
  x^M(-y)^N.
\end{equation}
Thus
\begin{equation}
  \Phi_x+3\Phi_y = \log\eta.	\label{eq:BJ2yo}
\end{equation}

Asymptotic behaviour tells us that
\begin{equation}
\eqalign{(B+iJ)_x&=\frac1{2\pi i} (\Phi_x+\pi i)\cr
	(B+iJ)_y&=\frac1{2\pi i} \Phi_y.\cr}
\end{equation}


\section{Interpretation}	\label{s:int}

Now that we have mapped out the values of $B+iJ$ around a region of
moduli space where some potentially small distances appear let us look
at the results. It will be simpler first to discuss the second example
of section \ref{s:hyp}, i.e., the blow-up of the $\Z_3$-quotient
singularity. First it is easy to reproduce the result of \cite{AGM:sd}
by inserting $\xi=\eta=0$ into (\ref{eq:BJ2xo}) to give
$\Phi_x=0$, i.e., $J_x=0$ and $B_x=\ff12$. We also have from
(\ref{eq:BJ2yo}) that $J_y=\infty$. This is interpreted as saying that
this limit point in the moduli space corresponds to a space where the
exceptional divisor (whose size is controlled by $J_x$) has zero size
but that another size elsewhere in the target space
(whose size is controlled by $J_y$) is infinite.

Although a zero size has appeared which one might consider unstringy,
one should have expected this result from (global) orbifold theory. Saying that
the exceptional divisor has zero size is like saying that the blow-up
has not been performed, that is we still have a quotient singularity.
We know from the arguments of \cite{W:phase} and \cite{AGM:II}
however that the point $\xi=\eta=0$ is precisely where the
large-radius orbifold conformal field theory should be. Thus our
analysis of $B+iJ$ fits in with the conformal field theory picture.
This can be stated in terms of the commutativity of the following
diagram
\begin{equation}
\def\mapright#1{\smash{\mathop{\longrightarrow}\limits^
	{\hbox{\scriptsize #1}}}}
\def\mapdown#1{\Big\downarrow\rlap{$\vcenter{\hbox{\scriptsize #1}}$}}
\def\arraystretch{1.5}
\begin{array}{ccc}
	M&\mapright{\sm}&S_M\\
	\mapdown{/G}&&\mapdown{/G}\\
	M/G&\mapright{``\sm''}&(S_M)/G
\end{array}
		\label{eq:com1}
\end{equation}
In the above diagram $M$ represents a smooth \CY\ manifold. This is
associated with a conformal field theory $S_M$ via the \nlsm. The
manifold admits a holonomy preserving symmetry $G$ which can be modded
out by to yield the orbifold $M/G$. The conformal field theory $S_M$
may also be modded out directly by adding in twisted sectors and
then projecting out all non-$G$-invariant states. One may also apply
the \nlsm\ to the orbifold $M/G$ by including the concept of twisted
strings in the usual \sm\ picture \cite{DHVW:}. The latter
modification is the reason for the quotation marks around ``\sm'' in
the diagram. Our statement about zero sized exceptional divisor may
then be viewed in terms of the commutivity of (\ref{eq:com1}). Since
$\eta=0$ may be considered as the large radius limit for $M$, each
link in the diagram should be well-defined and thus the diagram should
indeed commute. Let us now address the question of commutivity when $M$
is not at the large radius limit.

The form of (\ref{eq:BJ2xo}) and the cuts (\ref{eq:xer}) tell us that,
at least for small $\xi$ and $\eta$, we have $J_x\geq0$ with equality
only achieved for $\xi=\eta=0$. Thus $J_y$ {\em must\/} be infinite to
allow zero sized exceptional divisor. For a given small, nonzero
$\eta$ we have
\begin{equation}
  \eqalign{J_x&\geq-\frac{3\Gamma(\ff23)}{2\pi\Gamma^2(\ff13)}
	\hbox{Re$(\eta^{\frac13})=0$}\cr
	J_y&\simeq-\frac1{6\pi}\log|\eta|.\cr}
\end{equation}
Thus for a given value of $J_y$, i.e., $|\eta|$, we achieve minimum
$J_x$ by hugging the branch cuts. As discussed in section \ref{s:hyp},
the branch cut structure often warps as one moves out from the limit
point but to leading order we may assume the cuts are fixed. Thus we
may put $\arg(\eta)=2\pi$ in the above to obtain the asymptotic form
of the limit on $J_x$. Doing this we obtain
\begin{equation}
  J_x\geq\frac{3\Gamma(\ff23)}{4\pi\Gamma^2(\ff13)}e^{-2\pi J_y},
	\qquad{\rm for\ } J_y\gg1.	\label{eq:nlKc}
\end{equation}
This result embodies the true non-linear structure of the quantum
K\"ahler cone of string theory. Indeed, one might argue that the term
``cone'' should be dropped in view of the non-linearity of
(\ref{eq:nlKc}).

We may now address the commutivity of the diagram (\ref{eq:com1}) in the
case of a non-trivial $M$. The case we are looking at is not a global
orbifold but, as explained earlier, this is not expected to be
significant. The question we need to ask concerns the location of
points in $\cMc$ that correspond to the orbifolded conformal field
theory. We claim that these are naturally identified with the locus
$\xi=0$ at least in the neighbourhood of $\xi=\eta=0$ because of the
order 3 monodromy of the periods around this curve in $\cMc$. These
then correspond to the minimum size $J_x$ may obtain for a given $\eta$.
If $\eta\neq0$
this size is nonzero. Thus (\ref{eq:com1}) does not then commute. That is,
if we go from $M$ to $M/G$ classically, naturally the exceptional
divisor has zero size. If instead we go clockwise around the diagram
via $S_M$ and $(S_M)/G$ we arrive at nonzero size for nonzero $\eta$.

One might worry that (\ref{eq:nlKc}) suggests some large scale
non-local effects in the target space. If we ignore issues such as the
dimension and signature of space-time we might consider the scenario
in which the universe is shaped like the \CY\ space in question. If
$J_y$ represents the size of the universe then we may measure this by
``measuring'' (in a way unspecified) the minimum size of $J_x$. Thus
one appears to make global observations concerning the size of the
universe by only studying a very small part of it. Aside from the
hopeless impracticality of such a suggestion --- putting the order of
the size of the observable universe in $J_y$ we would have to measure
areas of order $10^{-(10^{100})}\alpha^\prime$ for $J_x$ --- this
reasoning turns out to be flawed for reasons we will explain below.

Consider the neighbourhood of an isolated quotient singularity as we
vary the sizes of the the rest of the target space. In the
large-radius limit of rest of the space, the neighbourhood of the
quotient singularity will become flat except for the singularity
itself --- i.e., the space will be locally {\em isometric\/} to
$\C^d/G$ for some $G$. Direct analysis of $\C^d/G$ always puts the
orbifold point in the ``right place'', i.e., the exceptional divisor
is zero size for the orbifold \cite{AGM:sd,me:orb2}. A better
hypothesis at this point might then seem to
be that exceptional divisors can only be shrunk down to zero size if
the neighbourhood around the resultant singularity becomes flat in the
process. The solves the apparent non-locality in the experiment to
measure the size of the universe. What we were measuring would not
actually be the size of the universe but the flatness of the space
surrounding the exceptional divisor. It would be the assumptions about the
global geometry of the universe which allowed us to infer its size.

An interesting case is that in which $M$ is a torus in
(\ref{eq:com1}). In this case, the neighbourhoods of the quotient
singularities are always flat independent of the size of the original
torus. It would be interesting to check explicitly to see if zero-sized
exceptional set can be achieved for these cases. Unfortunately the
construction of \cite{Bat:m} does not cover this case. One may be able
to use the construction of \cite{Boris:m} however. One might also
argue that when $M$ is a torus, each link in (\ref{eq:com1}) is
somehow exact and thus the diagram should commute.

Let us now look at the first example in section \ref{s:hyp} in which a
$\Z_2$-quotient singularity was analyzed. First recall that in
\cite{AGM:sd} it was shown that the $\Z_2$-quotient singularity
blow-up mode had the following form
\begin{equation}
  (B+iJ)_l = \frac1{\pi i}\cosh^{-1}z_l^{-\frac12},
\end{equation}
where $(B+iJ)_l$ is the component of the K\"ahler form measuring the
size of the blow-up, all other components of the K\"ahler forms are
taken to be infinite and $z_l$ is the ``algebraic'' coordinate in
$\cMc$ assumed to be in the form (\ref{eq:z2a}) so that
$z_l=0$ is the large blow-up limit,
$z_l=\infty$ in the orbifold and normalized so that
$z_l=1$ lies on the discriminant. (To be more precise on this latter
point, the whole rational curve given by varying $z_l$ including the
point at infinity may lie in the discriminant. In this case another
component of the discriminant will intersect this curve at $z_l=1$.)
Note that $J_l=0$ for $z_l=1$. This is equivalent to saying that
$J_l=0$ lies on the boundary of the \CY\ phase. Thus skeptics who
might suggest that we are only measuring zero distances in
\cite{AGM:sd} and this paper because of our definition of distance in
terms of analytical continuation should see that, in the
$\Z_2$-quotient singularity case, zero distances appear without the
need for analytical continuation. The special behaviour of the $\Z_2$
case can be seen by looking at the neighbourhood of the orbifold point
in $\cMc$. In general, a blow-up mode $(B+iJ)_l$ will have finite
monodromy around the orbifold point whose order we denote
$K_l$ . For the examples
in this paper, $K_l=2$ for the $\Z_2$-quotient singularity and
$K_l=3$ for the $\Z_3$-quotient singularity. More generally there will
be more than one blow-up associated to a single singularity and the
situation is more complicated. Still
considering for the moment the case where all other components of the
K\"ahler form are held at infinity, near the orbifold point, the behaviour
is generally of the form \cite{me:orb2}:
\begin{equation}
  (B+iJ)_l = \ff12+Re^{\pi i(\frac1{K_l}+\frac12)}z_l^{-\frac1
  {K_l}} + \ldots,	\label{eq:BJgeno}
\end{equation}
where $R$ is some positive real number and $0<\arg(z_l)<2\pi$. Thus
$K_l>2$, $J_l=0\Rightarrow z_l=\infty$. That is, zero size only
appears at the orbifold point. For $K_l=2$ however, following the
branch cut out from the orbifold point maintains $J_l=0$. Thus, zero
size is easier to acquire in the case $K_l=2$.

Considering now the case where the other components of the K\"ahler
form are allowed finite values we see from (\ref{eq:BJx1o}) that
$(B+iJ)_x$ is of the
form (\ref{eq:BJgeno}) for $K_l=2$ for both coordinates in $\cMc$.
Thus the situation is similar to the above paragraph. The condition
$J_x=0$ may be maintained by following the branch cuts out from the
orbifold point in $\cMc$ --- in either the $\xi$ or the $\eta$
direction.
This tells us that we may maintain zero size
for the blow-up mode while having finite size for the rest of the \CY\
manifold. This situation in thus different from the above.
In the case of the first orbifold, to obtain zero size we
required flatness for the target space metric (except at the
singularities) but in the case given by non-zero $\eta$ for the first
example in section \ref{s:hyp} we do not.

In conclusion it appears difficult to make a clear statement about
measuring zero distances within \CY\ spaces. There is evidence
\cite{AGM:sd} that the whole space can never be shrunk to zero. Indeed
any shrinking down which would lower the dimension of the target space
appears to be ruled out. For the case of blowing-down an exception
divisor to zero size, a sufficiently complicated topology will demand
that some other part of the target space must become infinitely large.
However, in simple cases where such a blow-down results in flatness or
$K_l=2$, this latter condition appears not to be necessary.


\section*{Acknowledgements}

It is a pleasure to thank P. Candelas, B. Greene, D. Morrison, R. Plesser
and E. Witten for useful conversations.
The author was supported by an NSF grant PHYS92-45317.

\end{document}